\documentclass[11pt]{article}
\usepackage{latexsym}
\usepackage{amsmath}
\usepackage{amssymb}
\usepackage[dvipdfmx]{graphics,color}
\usepackage[dviout]{graphicx}
\usepackage[dvipdfmx]{graphics}

\begin{document}

\title{Gradient flow model of mode-III fracture in Maxwell-type viscoelastic materials}
\author{Yoshimi Tanaka , Takeshi Takaishi}
%\email{taketaka@musashino-u.ac.jp}
%\affiliation{Musashino University, 3-3-3 Ariake, Koto-ku, Tokyo 135-8181, Japan}

%\author{Yoshimi Tanaka}
%\email{tanaka-yoshimi-vm@ynu.ac.jp}
%\affiliation{Graduate School of Environment and Information Science,
%Yokohama National University, 79-7 Tokiwadai, Hodogaya-ku, Yokohama 240-8501, Japan}
\date{\today}% It is always \today, today,
% but any date may be explicitly specified
%\keywords{Maxwell material, Phase field, Crack growth, Numerical simulation}

\maketitle

\begin{abstract}
We formulate a phase field crack growth model
for mode III fracture in a Maxwell-type viscoelastic material.
To describe viscoelastic relaxation, a field variable
of viscously flowed strain is employed in addition to a displacement field and damage phase field used in the original elastic model. Unlike preceding models constructed in the mechanical engineering community,
our model is based only on the generic procedure for driving
(uni-directional) gradient flow system
from a physically natural system energy
and employ no additional assumption
such as the super-imposed relations for stress and strain
(and their time derivatives) valid only for linear viscoelasticity.
Numerical simulations indicate that the competition
between increase in deformation by applied loading and
the viscoelastic relaxation determines whether a
distinct crack propagation has occurred
from an initial crack. Furthermore, we consider the numerical results from an energetic perspective.

\end{abstract}

\section{Introduction}
Fracture in soft polymeric materials is a subject
closely related to modern industrial society
and our daily life~\cite{ashby2018materials}.
The endurance of tires is, for example, crucial
for the safety of transportation infrastructures.
In machine assembly, controlling the toughness of adhesive,
i.e. weakly crosslinked polymers in rubbery state, is essential.
In the food industry,
comfortably and satisfyingly breaking polymers in paste or gel states is required \cite{van2013rheology}.

The fracture of soft polymers involves the viscoelastic effect.
Thus, its analysis is outside the applicability
of the typical linear elastic fracture mechanics~\cite{anderson2017fracture}.
Thus far, several experiments have been performed regarding viscoelastic effects on the fracture of soft polymers~\cite{gent1994viscoelastic,knauss2015review, creton2016fracture, poulain2018damage},
and a variety of analytical and numerical models
(primarily on the continuum scale)
have been proposed to understand the experimental results
\cite{schapery1975theory, hui1992fracture, de1996soft, rahulkumar2000cohesive, persson2005crack}.
With respect to the dynamical aspect, however, most theoretical models
are specialised, that is, they intend to describe the fracture
behaviour of specific materials in specific conditions
(e.g. steady-state crack propagation in rubbers).
Meanwhile,
to broaden our understanding on viscoelastic fracture,
a general
and flexible theoretical framework
that can compile the experimental
and theoretical results obtained thus far is required,
similar to the time-dependent Ginzburg-Landau
theory for phase transition dynamics~\cite{onuki2002phase}.

A potential candidate for such a framework
is the phase field fracture model (PFFM)
that was first proposed to describe
quasi-brittle crack propagations
in isotropic linear elastic materials~\cite{aranson2000continuum, bourdin2000numerical, karma2001phase}
and subsequently extended to various materials
with complex mechanics properties~\cite{ambati2015phase, miehe2015phase,carrara2017consistent,chukwudozie2019variational} including
viscoelastic materials~\cite{miehe2014phase, shen2019fracture}.
The PFFM introduces a field variable (damage phase field)
representing the extent of damage (reduction in elasticity),
and treats crack propagations as growth of
a fully damaged narrow {\it domain} (with zero elastic modulus),
instead of as the formation of a new crack {\it surface}.
Hence,
we can avoid the difficulty of free-boundary problems
involving the motion of the crack tip that is a singular point of a stress field
in typical continuum description.
A key point of the currently interested PFFM is
the employment of regularized system energy
which coincides with the Griffith energy
in the sharp crack limit~\cite{ambrosio1992approximation}.
By combining the regularized energy and additional assumptions
on time evolution (employing gradient-flow system or
almost alternative dissipative function), 
we can formulate fracture dynamics exhibiting high consistency
with the classical Griffith theory.

For example, one of the present authors (T.T.) and Kimura
formulated the dynamics of mode-III fracture
in linear elastic materials by considering
the gradient-flow system of a regularised system energy
with respect to the deformation
and the damage phase field~\cite{takaishi2009phase, takaishi2009numerical}.
Their numerical results indicated that
the derived time evolution equations
(the gradient-flow system) could predict
reasonable crack paths for complex boundary conditions.

This study aims to extend the gradient-flow PFFM 
for linear elastic materials in~\cite{ takaishi2009phase}
to the mode-III fracture in a Maxwell-type
viscoelastic material.
An important basis of this extension is the fact
that a certain class of viscoelastic constitutive relation
can be described as a gradient-flow system~\cite{kimura2019gradient}.
To describe viscoelastic stress relaxation,
a viscous strain is employed with the deformation
field in the expression of elastic energy;
a viscous strain represents the portion of the total strain flowed
in a viscous manner, corresponding to a dashpot's deformation
in symbolic one-dimensional (1d) rheological models; see Fig. \ref{fig:F1}.

The gradient-flow system of the energy
accords with the PDEs representing constitutive relations
and the force balance condition. The two gradient-flow
descriptions above (for crack propagation and for viscoelastic relaxation)
are combined to formulate the dynamics of viscoelastic fracture.
That is, we construct a system energy
in terms of the out-of-plane displacement field,
viscous strain field, and damage phase field,
and derive a gradient-flow system
as a set of time evolution equations of
the three field quantities.

It is emphasised that our formulation does not rely on
the superimposition principle for the strain rate and stress
(and the resultant the convolutional stress-strain
relation with the exponential kernel),
which is distinctive to the linear viscoelasticity.
This is the essential difference from
the previous study~\cite{shen2019fracture}.
Our formulation follows the typical mathematical
procedure to derive the gradient system,
that is, constructing the system energy
with a clear physical meaning and deriving its gradient system
to obtain a set of PDEs describing the time evolution of the system.
It could, in principle, include
geometrical and material nonlinearities
in rheological constitutive relations,
because of the independence of the superimposition principle.

Based on the obtained time evolution equations,
we performed numerical simulations
on a system with an initial crack subjected
to mode-III (anti-plane) loading
by boundary displacement with a constant velocity.
The numerical result indicates that
a distinct crack growth can occur from the initial crack
in a quasi-brittle manner if the loading velocity is sufficiently large.
Meanwhile, for a sufficiently slow loading, the system demonstrates ductility, that is,
the initial notch opens progressively
without extension.
The crossover between ductile/(quasi-)brittle behaviours
is caused by the competition between the increase in
deformation by the boundary displacement
and the viscoelastic relaxation.
This is qualitatively consistent with experiments
on some Maxwell-type viscoelastic
liquids~\cite{gladden2007motion, tabuteau2011propagation, huang2017polymer},
suggesting that the present simple model
(describing an artificial situation of a mode-III
fracture of a Maxwell-type material) can be a prototype
for more detailed modelling.
Additionally, we discuss our viscoelastic PFFM
from an energetic viewpoint.

\section{Model}
%Fig. 1%%%%%%%%%%%%%%%
\begin{figure}[]
\begin{center}
\includegraphics[width=0.7\linewidth]{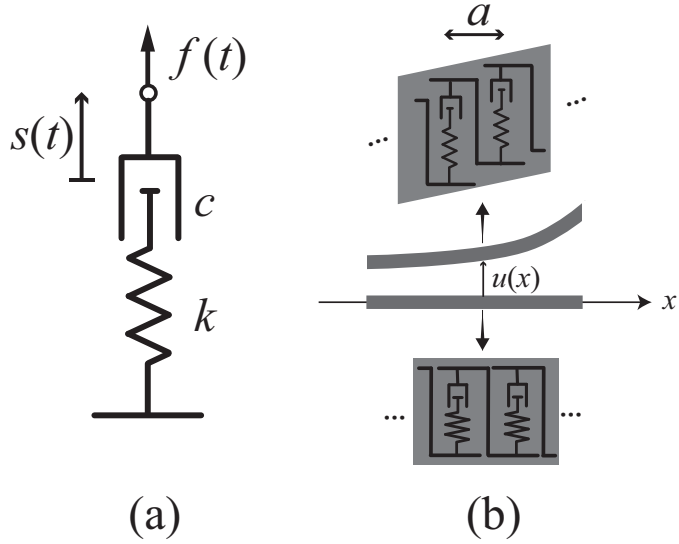}
\caption{(a) Maxwell element with stretching deformation $s(t)$ and tension $f(t)$.
(b) 1d continuum system comprising
an array of innumerable vertically oriented Maxwell elements.}
\label{fig:F1}
\end{center}
\end{figure}
%%%%%%%%%%%%%%%%

In this section, we explain the basic concept for describing crack propagation in viscoelastic materials as a gradient-flow system.
We begin by reconsidering the mechanical behaviour of a single Maxwell
element comprising a linear spring with elastic modulus $k$ and a dashpot
with viscous friction constant $c$ (Fig.~\ref{fig:F1}).
The tension $f(t)$ acting on the element and stretching deformation $s(t)$
satisfies $ f =k s_{\mathrm{e}} = c \dot{s}_{\mathrm{v}}$ and
$s=s_{\mathrm{e}} + s_{\mathrm{v}}$, where $s_{\mathrm{e}}$
and $s_{\mathrm{v}}$
are the elastic deformation of the spring and the ``viscously flowed''
deformation
of the dashpot, respectively;
the dot symbol `` $\dot{\, }$ '' 
represents the time derivative.
Eliminating $s_{\mathrm{e}}$ and $s_{\mathrm{v}}$ yields
%%%%%%%%%%%%%%%%%%%%%%%%
\begin{equation}
\label{eq:fk_cs}
\dot{f} + \left( \frac{k}{c} \right) f = c \dot{s}\, ,
\end{equation}
%%%%%%%%%%%%%%%%%%%%%%%%%
and eliminating $f $ and $s_{\mathrm{e}}$ yields
%%%%%%%%%%%%%%%%%%%
\begin{equation}
\label{eq:cs_ss}
c \dot{s}_{\mathrm{v}} = k \left( s- s_{\mathrm{v}} \right).
\end{equation}
%%%%%%%%%%%%%%%%
Eq. \eqref{eq:fk_cs} is the typical force--deformation relation
of a single Maxwell element,
indicating that the time derivative of force contains 
two contributions of the decay term with a characteristic relaxation time $\tau = c /k$
and the generation term proportional to the stretching rate $\dot{s}$.
Eq. \eqref{eq:cs_ss} indicate that the flowed component of the stretching deformation
$s_{\mathrm{v}}(t)$ relaxes toward the target value $s(t)$
with the relaxation time $\tau$.
It is noteworthy that $\left( s- s_{\mathrm{v}} \right)$ in the r.h.s. of eq. \eqref{eq:cs_ss}
is the elastic deformation.

Next, we discuss simple 1d continuum problem,
as shown in Fig. \ref{fig:F1}(b). The system consists of innumerous
identical Maxwell elements aligned along the $x$-axis with spacing $a$.
The deformations of the elements are confined in the vertical direction.
In the continuum description (the limit of $a \to 0$), the mechanical state of the system is described
by the out-of-line displacement $u(x, t)$ and the viscous strain $e(x, t)$,
i.e. the counterpart of $s_{\mathrm{v}}(t)$ of the ``microscopic'' Maxwell elements.
The time evolution of the system is determined by the constitutive relation
(corresponding with eq. \eqref{eq:cs_ss}),
%%%%%%%%%%%%%%%%%%%%%%%
\begin{equation}
\label{eq:he_te}
\eta_{\rm 1d} \dot{e} = \mu_{\rm 1d} \left( \partial_{x} u(x,t)-e \right),\end{equation}
%%%%%%%%%%%%%%%%%%%%
and the motion equation,
%%%%%%%%%%%%%%%%%%%%%%
\begin{equation}
\label{eq:ru_te}
\rho \ddot{u} = - \zeta_{\rm 1d} \dot{u} + \partial_{x} \mu_{\rm 1d} \left( \partial_{x}u\left( x,t \right)-e \right) \,
\end{equation}
%%%%%%%%%%%%%%%%%%%%%%%%%%%%
where $\mu_{\rm 1d}$ ($=k a$), $\eta_{\rm 1d}$ ($=c a$), $\rho$,
and $\zeta_{\rm 1d}$
are the 1d shear modulus, 1d viscosity, 1d mass density, and 1d viscous friction coefficient for the background, respectively.
When the inertia term is negligible,
eqs. \eqref{eq:he_te} and \eqref{eq:ru_te} become a set of gradient-flow equations of
%%%%%%%%%%%%%%%%%%%%%%%
\begin{equation}
\label{eq:zu_du}
\zeta_{\rm 1d} \dot{u} = - \frac{{\delta E}_{\rm{1d}}}{\delta u}
\end{equation}
%%%%%%%%%%%%%%%%%%%%%%
\begin{equation}
\label{eq:he_de}
\eta_{\rm 1d} \dot{e} = -\frac{{\delta E}_{\rm{1d} } }{\delta e},
\end{equation}
%%%%%%%%%%%%%%%%%%%%%%%%
where
%%%%%%%%%%%%%%%%%%%%
\begin{equation}
\label{eq:E_ue_}
E_{\mathrm{1d}}:=\frac{\mu_{\rm 1d} }{2} \int \mathrm{d}x
\left( \partial_{x}u-e \right)^{2}\ 
\end{equation}
%%%%%%%%%%%%%%%%%%%%
is the elastic (deformation) energy of this 1d model.

%%%%%%%%%%%%%%%%
% Figure 2
\begin{figure}[]
\begin{center}
\includegraphics[width=0.9\linewidth]{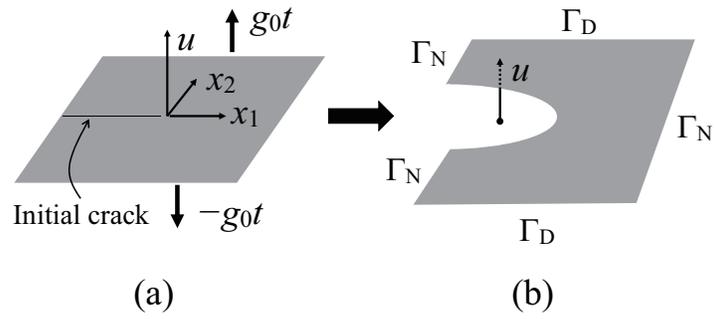}
\caption{Computational domain in our simulation at (a) the initial state
and (b) a state with an opened crack. $\Gamma_{\mathrm{D}}$ and
$\Gamma_{\mathrm{N}}$ represent the Dirichlet and Neumann boundaries, respectively.}\label{fig:F2}
\end{center}
\end{figure}
%%%%%%%%%%%%%%%%
To construct a PFFM for a Maxwell-type material under
mode-III loading,
we consider a two-dimensional (2d)
system undergoing an out-of-plane displacement
$u(\mathrm{\mathbf{x}},t)$,
where $\mathrm{\mathbf{x}}=(x_{1}, x_{2})$ is a position vector
indicating a point on the system (see Fig.~\ref{fig:F2}).
Furthermore, we introduce a phase field $z(\mathrm{\mathbf{x}}, t)$
representing the damage of the 
system such that $z =0$ corresponds
to a non-damaged state and $z =1$
corresponds to a completely broken state.

We assume that the total energy $E$ of the system consists
of the deformation energy $E_{\mathrm{def}}$ and the damaging energy $E_{\mathrm{dam}}$:
%%%%%%%%%%%%%%%%%%%%%%%%%
\begin{equation}
\label{eq:E_EE}
E=E_{\mathrm{def}}+E_{\mathrm{dam}}\, , 
\end{equation}
%%%%%%%%%%%%%%%%%%%%%%%%%%%
\begin{equation}
\label{eq:E_ue_2}
E_{\mathrm{def}}:=
\int_\Omega {\mathrm{d\mathbf{x}}\,
{\left( 1-z \right)^{2} \frac{\mu }{2} \left( \mathrm{\nabla }u-\mathrm{\mathbf{e}} \right)}^{2}}, 
\end{equation}
%%%%%%%%%%%%%%%%%%%%%%%%%%%%
\begin{equation}
\label{eq:E_ze}
E_{\mathrm{dam}}:= \frac{\gamma }{2} \int_\mathrm{\Omega } 
{\mathrm{d\mathbf{x}}\, \left( \epsilon \left| \mathrm{\nabla }z 
\right|^{2}+\frac{z^{2}}{\epsilon } \right)},
\end{equation}
%%%%%%%%%%%%%%%%%%%%
where $\Omega$ represents a 2d domain on which the model
is defined, and the constants $\mu$ (N$\cdot$m$^{-1}$),
$\gamma$ (N), and $\epsilon$ (m) are the
2d shear modulus, 2d fracture surface energy,
and a length scale regulating the singular behaviour of the deformation field,
respectively ($\epsilon$ is the thickness of a diffusive fracture surface).
The integral in eq. \eqref{eq:E_ue_2} measures the length of the crack
in the purely elastic model~\cite{takaishi2009phase}.
It is noteworthy that in the out-of-plane problem,
the strain becomes a vector quantity $\nabla u$
(where $\nabla =(\partial / \partial x_1, \partial / \partial x_2)$);
correspondingly, we introduce the viscous strain vector 
of $\mathrm{\mathbf{e}}=(e_{1},\, \, e_{2}\mathrm{)}$.

We employ the gradient-flow system of $E$ as a set of time evolution
equations:
%%%%%%%%%%%%%%%%%%%%%%%%
\begin{equation}
\label{eq:au_ue}
\alpha _{u} \dot{u} = -\frac{ \delta E }{ \delta u }
= \mu \ \mathrm{div} \left( (1-z)^{2} (\nabla u - \mathbf{e}) \right)
\end{equation}
%%%%%%%%%%%%%%%%
\begin{equation}
\label{eq:ae_ue}
\alpha _{e} \dot{\mathbf{e}}=-\frac{ \delta E }{\delta \mathbf{e}} =\mu \, ( 1-z )^{2} ( \nabla u - \mathbf{e}) 
\end{equation}
$$
\left( \alpha _{e} \dot{e}_{i}=-\frac{ \delta E}{\delta e_{i}}
=\mu \, ( 1-z )^{2} ( \partial_{x_{i}}u - e_{i}) \,\,\,\, i=1,2 \, \right)
$$
%%%%%%%%%%%%%%%%%%%%%
\begin{equation}
\label{eq:az_ez}
\alpha_{z} \dot{z} = \left( - \frac{\delta E } {\delta z} \right)_{+}
=\left( \gamma \left( \epsilon \Delta z-\frac{z}{\epsilon } \right)
+ \mu \left| \nabla u - \mathbf{e} \right|^{2} ( 1-z ) \right)_{+},
\end{equation}
%%%%%%%%%%%%%%%%%%%%%%%%%%%%
where the non-negative constants
$\mathrm{\alpha}_u$, $\mathrm{\alpha}_e$,
$\mathrm{\alpha}_z$ are
kinetic coefficients; $\dot{A}:=\partial A/\partial t$
and the operation $ ( \cdot )_{+} := {\rm max}(\cdot, 0) $
ensure the irreversibility of crack growth.

The following boundary conditions are imposed for
$u(\mathrm{\textbf{x}},t)$ and $z(\mathrm{\textbf{x}},t)$:
%%%%%%%%%%%%%%%%%%%
\begin{equation}
\label{eq:u_xt}
u=g\left( \mathrm{\mathbf{x}},t \right)\, \, \, \mathrm{\mathbf{x}}\in 
\Gamma_{ \mathrm{D} }\, ,
\end{equation}
%%%%%%%%%%%%%%%%%%%%%%%%%
\begin{equation}
\label{eq:nu_}
\left( \mathrm{\nabla }u-\mathrm{\mathbf{e}} \right)\cdot \mathrm{\mathbf{n}}=0 \,\,\, \mathrm{\mathbf{x}} \in \Gamma_{\mathrm{N}} \, ,
\end{equation}
%%%%%%%%%%%%%%%%%%%%%
\begin{equation}
\label{eq:dz_}
\frac{\partial z}{\partial n}=0\, \, \,\,\, \mathrm{\mathbf{x}}\in 
\Gamma,
\end{equation}
%%%%%%%%%%%%%%%%%
where $g\left( \mathrm{\mathbf{x}},t \right)$ is a given function,
$\Gamma $ is the boundary of $\Omega $, and $\Gamma_{\mathrm{D} }$ and $\Gamma_{\mathrm{N} }$ are the parts of $\Gamma$
($\Gamma = \Gamma_{\mathrm{D}} \cup \Gamma_{\mathrm{N} } $) 
with Dirichlet and Neumann boundary conditions, respectively.
The time dependence of $g(\mathbf{x}, t)$ implies that
a protocol of boundary loading is to be prescribed.
Eq. \eqref{eq:nu_} represents the stress-free condition
of the boundary.
With adequate initial conditions of $u$ and $z$,
eqs. \eqref{eq:au_ue}--\eqref{eq:az_ez} and eqs. \eqref{eq:u_xt}--\eqref{eq:dz_}
determine the time evolution of the system.

From the correspondences between eqs. \eqref{eq:he_te}--\eqref{eq:ru_te}
and eqs. \eqref{eq:au_ue}--\eqref{eq:ae_ue},
the constants $\alpha_{u}$ and $\alpha_{e}$ can be
identified as the friction with the background and internal viscosity,
respectively.
We hence denote $\alpha_{\rm e}$ as $\eta$.
Although $\alpha_{z}$ is of the same dimension as
$\alpha_{\rm e}$ (i.e. $\eta$),
its physical meaning is different:
$\alpha_{z}$ can be related to
the crack velocity dependence of the effective fracture energy.
The details for the interpretation of $\alpha_{z}$ will be reported elsewhere.
Below, we treat a special case of
%%%%%%%%%%%%%%
\begin{equation}
\label{eq:a_}
\alpha_{u}=0
\end{equation}
%%%%%%%%%%%%%%
and
%%%%%%
\begin{equation}
\label{eq:a_a}
\alpha_{e} \gg \alpha_{z}.
\end{equation}
%%%%%%%%%%%%%%

To investigate the energy balance of the present model, we take the time derivative of the system energy, $\dot{E} =dE/dt=\int_{\Omega} dx dy \left( \frac{\delta E}{\delta u} 
\dot{u}
+ \frac{\delta E}{\delta \mathbf{e}} \dot{\mathbf{e}}
+\frac{\delta E}{\delta z} \dot{z} 
\right) + \mathrm{B. T.} $, where B. T. means the boundary terms.
With Eqs. \eqref{eq:au_ue}--\eqref{eq:az_ez}, zero Neumann boundary
conditions for $u$ and $z$ (\eqref{eq:nu_} and \eqref{eq:dz_}) and the
additional conditions of \eqref{eq:a_} and \eqref{eq:a_a} (that is, $\delta E/ \delta u =\alpha_{u} \dot{u} =0 $ and $ \left( \delta E/ \delta z \right)_{+} = \alpha_{z} \dot{z} \approx 0$ ), we have
%%%%%%%%%%%%%%
\begin{equation}
\label{eq:E_WQ}
\dot{E}=\dot{W}-\dot{Q}
\end{equation}
%%%%%%%%%%%%%%%%%%%%%%%
where 
%%%%%%%%%%%%%
\begin{equation}
\label{eq:W_en}
\dot{W}=\int_{\Gamma_{\mathrm{D}}} dc\, \dot{u} \mu (1-z)^2 ( \nabla u - \mathbf{e}) \cdot \mathbf{n}
\end{equation}
%%%%%%%%%%%%%%%%%%
and
%%%%%%%%%%%%%%%%%%%%%%%%%%%%
\begin{equation}
\label{eq:Q_ee}
\dot{Q}=\int_{\Omega} dx dy\, \mu (1-z)^2 (\nabla u - \mathbf{e}) \cdot \dot{\mathbf{e}}
\end{equation}
%%%%%%%%%%%%%%%%%%%%
are the work and heart generation per
unit time of the entire system, respectively.

Eq. \eqref{eq:E_WQ} or $\dot{W}=\dot{E}+\dot{Q}$
indicates that the work performed to the system
is partitioned to system energy and
heat generation.

\section{Numerical results}
The system is a square domain $\Omega$ with side length $L$
(i.e. $ \mathbf{x} =(x_1, x_2) \in \Omega = (-\frac{L}{2}, \frac{L}{2})^2 $).
An initial crack (highly damaged narrow zone) is prepared along the negative part of the $x_{1}$-axis
(see Fig. \ref{fig:F2}). The system is loaded
by anti-symmetrical displacements on $\Gamma_{\mathrm{D}}$, $ u(\mathbf{x},t)|_{ x_{2}=\pm L/2} = \pm g_{0} t$,
where $g_{0} $ is the speed (m$\cdot$s$^{-1}$)
of the boundary displacements.

For numerical simulations, we non-dimensionalise
eqs. \eqref{eq:au_ue}--\eqref{eq:az_ez} by employing
$L$, $T:=L/g_{0}$ and $L \mu$ as the units
of length, time, and force, respectively.
Introducing the non-dimensionalised quantities of
$\tilde{x}_i =x_{i}/L$ 
($\partial/\partial x_i=L^{-1}\partial/\partial \tilde{x}_i$),
$\tilde{u} =u/L$, $\tilde{\varepsilon}= \varepsilon/L$,
$ \tilde{t}=t/T$ ($\partial/\partial t=T^{-1}\partial/\partial \tilde{t}$),
$ \tilde{\gamma}=\gamma/(L \mu)$,
$ \tilde{\alpha}_z=\alpha_z/(T \mu)$
and
%%%%%%
\begin{equation}
\label{eq:g_Tm}
\tilde{g}_0 =\frac{\alpha_e (=\eta)}{T \mu}.
\end{equation}
%%%%%%
Eqs. \eqref{eq:au_ue} - \eqref{eq:az_ez} (with $\alpha_{u}=0$) become
%%%%%%%%%%%%%%%%%%%%%%%%
\begin{equation}
\label{eq:_ue}
0= \mathrm{div} \left( (1-z)^{2} (\nabla u - \mathbf{e}) \right)
\end{equation}
%%%%%%%%%%%%%%%%
\begin{equation}
\label{eq:ge_ue}
\tilde{g} _{0} \dot{\mathbf{e}}= ( 1-z )^{2} ( \nabla u - \mathbf{e}) 
\end{equation}
%%%%%%%%%%%%%%%%%%%%%
\begin{equation}
\label{eq:az_ez_2}
\tilde{\alpha}_{z} \dot{z} =\left( \tilde{\gamma} \left( \tilde{\epsilon} \Delta z-\frac{z}{\tilde{\epsilon} } \right)
+ \left| \nabla u - \mathbf{e} \right|^{2} ( 1-z ) \right)_{+},
\end{equation}
%%%%%%%%%%%%%%%%%%%%%%%%%%%%
where and hereinafter we omit the `` $\tilde{\ }$ '' over $u$ and
$\mathbf{x}$ and $t$, and `` $\dot{\ }$ '' and $\nabla$ etc.
represent the derivatives with respect to
the non-dimensionalised time and spatial coordinates.
(the tilde symbol for the model parameters of $\tilde{\gamma}$
$\tilde{\varepsilon}$ and $\tilde{g}_0 $ is preserved.)
Because we use $L$ as the unit of length,
$\Omega=(-1/2, 1/2)^2$ and the non-dimensionalised boundary
displacement is $g_{0}t/L =t/(L/g_{0})=\tilde{t}=t$
(the last equality merely represents the omission of the tilde symbol).
The boundary displacement velocity is always unity
in the present non-dimensionalisation.
Furthermore, the energetic quantities of $E$, $E_{\mathrm{def}}$,
$E_{\mathrm{dam}}$, $W$, and $Q$ hereinafter represent
the non-dimensionalised ones scaled by the characteristic energy
$L^2 \mu$.
According to the these abbreviations,
$\dot{E}$ represents ``$(\mu L^2/T)^{-1} dE/dt$''
in the original dimensional notation.

It is noteworthy that $\tilde{g} _{0} =\frac{g_0}{L/\tau}$ ($T=L/g_{0}$),
where $\tau=\eta/\mu$ is the viscoelastic relaxation time;
$\tilde{g}_{0}$ is the ratio between the loading velocity
and the characteristic velocity of $L/\tau$.

%%%%%%%%%%%%%%%%%%%%%%%%%%%%%%%%%%%
%Figure 3%%%%%%%%%%%%%%%%%%%%%%%%%%%
\begin{figure}[htbp]
\begin{center}
\includegraphics[scale=0.6]{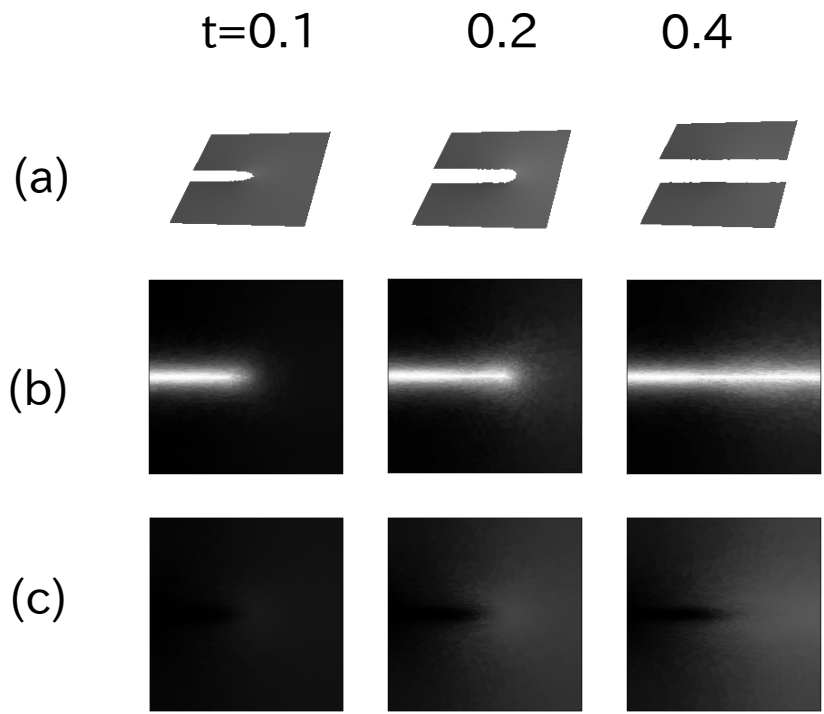}
\end{center}
\caption{ Temporal evolution of the three 
field variables of (a) deformation (bird-eye view),
(b) $z$, and (c) $| \mathbf{e} |$ (gray level mapping) for $\tilde{g}_0=0.1$. 
The bright zone in (b) represents a crack.}
\label{fig:Maxwell-mode3-mu1-gam-0.01-g0-1-alp3-0.1}
\end{figure}
% Figure 4%%%%%%%%%%%%%%%%%%%%%%%%
% alp3=0.02, g0=1
\begin{figure}[htbp]
\begin{center}
\includegraphics[scale=0.6]{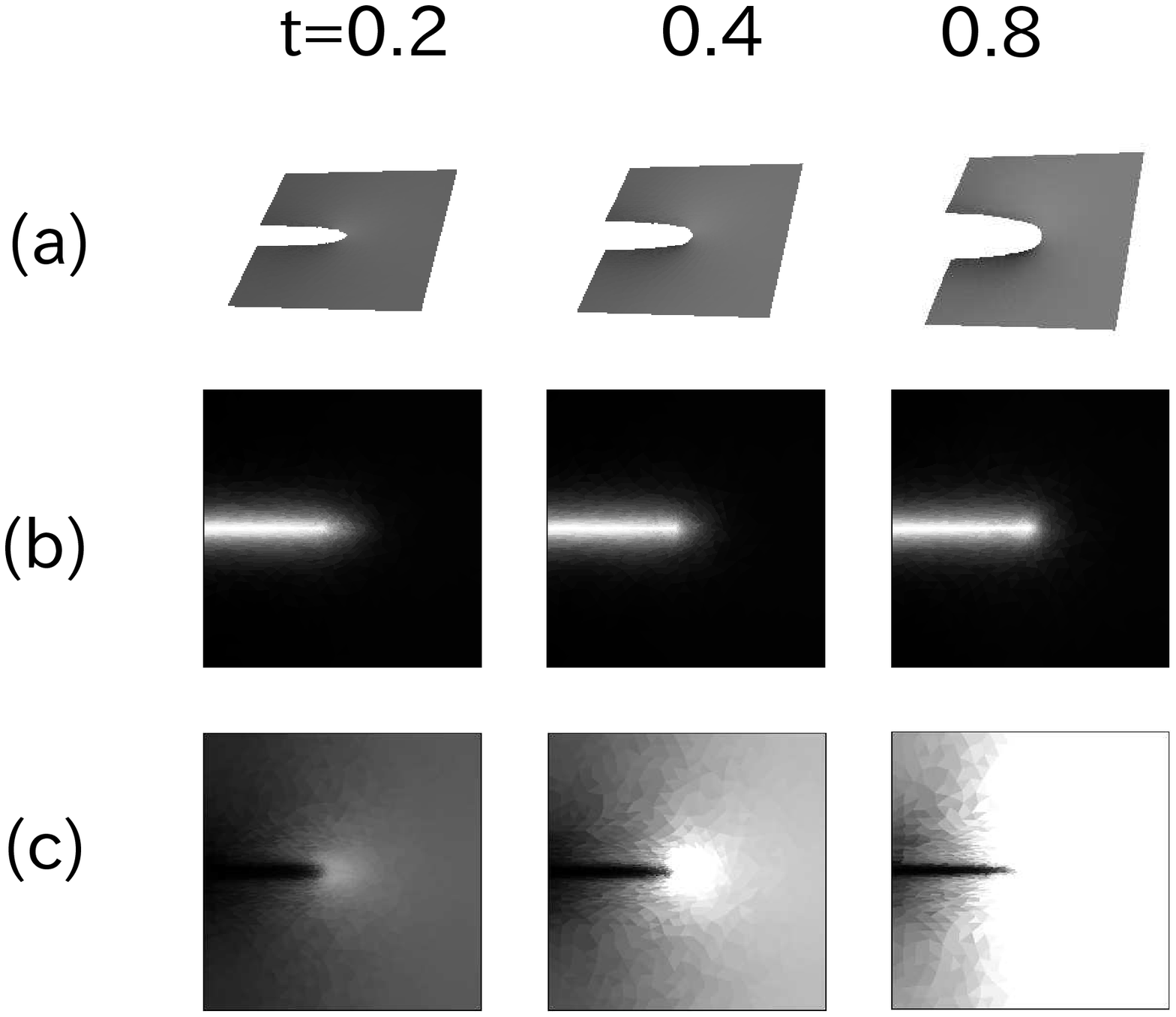}
\end{center}
\caption{Temporal evolution of the three 
field variables of (a) deformation (bird-eye view),
(b) $z$, and (c) $| \mathbf{e} |$ (gray level mapping)
for $\tilde{g}_0=0.02$. }
\label{fig:Maxwell-mode3-mu1-gam-0.01-g0-1-alp3-0.02}
\end{figure}
%%%%%%%%%%%%%%%%%%%%%%%%%%%
%%%%%%%%%%%%%%%%%%%%%%%%%%%%%
We set $\tilde{\gamma} = 0.01$ and $\tilde{\epsilon} = 0.05$,
and systematically change $\tilde{g}_0$ from 0.02 to 10.
The initial conditions are $u(\mathbf{x}, 0)=0$, $e_{i}(\mathbf{x}, 0)=0$ and $ z(\mathbf{x}, 0) =\mathrm{exp}(-x_{2}^{\ 2}/10^{-4}) / \left( 1+\mathrm{exp}(x_{1}/(10^{\-2}) \right) $; the third relation represents the initial crack.
Actual numerical simulations were performed with
FreeFem++~\cite{hecht2012new}.

Figures \ref{fig:Maxwell-mode3-mu1-gam-0.01-g0-1-alp3-0.1}
and \ref{fig:Maxwell-mode3-mu1-gam-0.01-g0-1-alp3-0.02}
exhibit the time evolution of the field variables
of the (a) displacement $u$, (b) damage phase field $z$
and (c) amplitude of the viscous strain vector $|\mathbf{e}|$
for different loading velocities of $\tilde{g}_{0}=0.1$ and 0.02.

At a relatively fast loading velocity of $\tilde{g}_{0}=0.1$,
the initial crack suddenly propagates when the boundary displacement
reaches a critical value.
At a slow loading of $\tilde{g}_{0}=0.02$,
the initial crack never extends;
instead, the amplitude of the viscous strain
progressively increases in the un-broken region
ahead of the initial crack tip with increasing the boundary displacement,
see the brightness (indicating larger $|\mathbf{e}|$)
in the rightmost figure
in Fig. \ref{fig:Maxwell-mode3-mu1-gam-0.01-g0-1-alp3-0.02}(c).
These two mechanical behaviour crossover at approximately
$\tilde{g}_{0}=0.04$.

\begin{figure}[htbp]
\begin{center}
\includegraphics[width=1\linewidth]{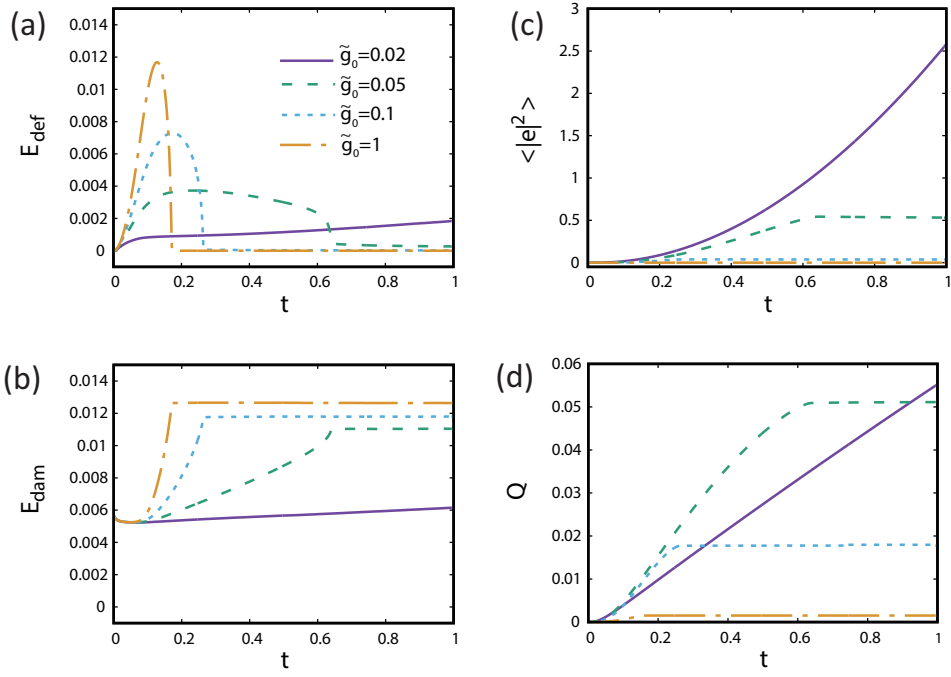}
\end{center}
\caption{Temporal evolution of
(a)deformation energy $E_{\textrm{def}}$,
(b)damaging energy $E_{\textrm{dam}}$,
(c) $\langle |\mathbf{e}|^2 \rangle$,
and (d) the total dissipation $Q$
for $\tilde{g}_{0}=0.02$ (solid line), $0.05$ (dashed line), $0.1$ (dotted line)
and $1$ (long dashed dotted line).}
\label{fig:Maxwell-mode3-g0-1-alp3-Eel-Es}
\end{figure}

Figures \ref{fig:Maxwell-mode3-g0-1-alp3-Eel-Es}(a)--(d) show $E_{\rm def}$, $E_{\rm dam}$, $\langle |\mathbf{e}|^2 \rangle :=\int_{\Omega}dx dy |\mathbf{e}|^2$
and dissipation $Q(t) :=\int_{0}^{t} \dot{Q}(s) ds$
(see eq. \eqref{eq:Q_ee} for the definition of $\dot{Q}$)
as functions of $t$ ($x$, $y$, and $t$ represent the non-dimensionalised quantities).
Each curve in Fig. \ref{fig:Maxwell-mode3-g0-1-alp3-Eel-Es}
shows the results for different $\tilde{g}_{0}$.
For the fast deformation of $\tilde{g}_{0}=1$,
$E_{\rm def}$ increases with $t$
to achieve the maximum and subsequently reduces to zero,
as shown in Fig. \ref{fig:Maxwell-mode3-g0-1-alp3-Eel-Es}(a).
$E_{\rm dam}$ starts to increase at the time ($t \approx 0.1$)
when $E_{\rm def}$ achieves the peak and reach its plateau
at the time ($ t \approx 0.18$) when $E_{\rm def}$ drops to zero
(the non-zero initial value of $E_{\rm dam}$
is due to $z( \mathbf{x}, 0)$ giving the initial crack).
.
This correlated behaviour of $E_{\rm def}$
and $E_{\rm dam}$ corresponds to a distinct (quasi-brittle)
crack growth 
Meanwhile, $ \langle |\mathbf{e}|^2 \rangle $
and $Q$ is almost zero for the fast loading
(Figs.~\ref{fig:Maxwell-mode3-g0-1-alp3-Eel-Es}(c) and (d)).

For the slow deformation of $\tilde{g}_{0}=0.02$,
the system exhibit ductility
(see also Fig. \ref{fig:Maxwell-mode3-mu1-gam-0.01-g0-1-alp3-0.1})
and the energetic quantities behave in the opposite manners:
$E_{\rm def}$ and $E_{\rm dam}$ remain on low levels
and $ \langle |\mathbf{e}|^2 \rangle $ increases progressively with $t$.

In the intermediate deformation velocity regime
($\tilde{g}_{0}=0.05$ and $0.1$),
$E_{\rm def}$ achieves a peak corresponding
to a distinct crack growth (and $E_{\rm dam}$ increases
correlatedly with the change of $E_{\rm def}$)
but the peak is dull and broaden.
The dissipation $Q$ in this intermediate regime
increases most rapidly until the crack growth is completed,
suggesting that the crack growth involves a large
amount of viscous dissipation. 

\begin{figure}[htbp]
\begin{center}
\includegraphics[width=1\linewidth]{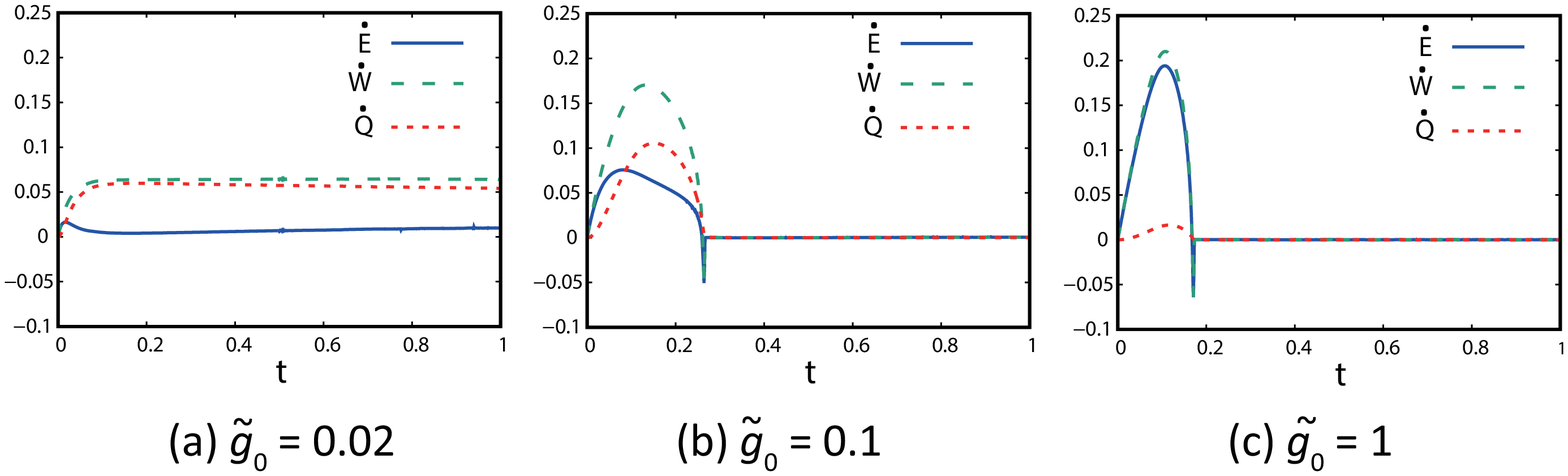}
\end{center}
\caption{Temporal evolution of $\dot{E}$(solid
line),
$\dot{W}$(dashed line) and $\dot{Q}$(dotted line)
when $\tilde{g}_{0}=0.02, 0.1$ and $1.0$.}
\label{fig:Maxwell-mode3-g02-1-dene}
\end{figure}

To view the energy balance of the system in detail,
we plot the time change of $\dot{W}$,
$\dot{E}$ ($=\dot{E}_{\rm def}+\dot{E}_{\rm dam}$),
and $\dot{Q}$ in Fig. \ref{fig:Maxwell-mode3-g02-1-dene}.
For the slow loading of (a) $\tilde{g}_{0}=0.02$,
$\dot{E}$ remains, after the initial transient period with
a small peak, low level (strictly, $\dot{E}$ slightly increases with $t$),
while $\dot{W}$ and $\dot{Q}$ increase rapidly in the initial period
and subsequently reach plateau regions
(the plateau of $\dot{Q}$ is slightly sloped, corresponding to the slight
increase in $\dot{E}_{\rm def}$).
After the appearance of the plateaus,
\[ \dot{W} \gtrsim \dot{Q} \gg \dot{E} \]
holds. Most of the work dissipate.

Meanwhile, for the fast loading of (c) $\tilde{g}_{0}=1$,
$\dot{W}$ and $\dot{E}$ achieve shape peaks and subsequently
reduce to zero, corresponding to
the distinct crack growth.
Before the crack growth is completed ($t \le 0.18$),
it holds that
\[ \dot{W} \gtrsim \dot{E} \gg \dot{Q}, \]
most of the work performed by the boundary
displacement transfers into increases in the system energy,
and the amount of viscous dissipation is small.
The system fractures in a quasi-brittle manner.

For the intermediate loading velocity
of (b) $\tilde{g}_{0}=0.1$,
$\dot{W}$ and $\dot{E}$ exhibit peaks
(that is, a distinct crack growth occurs),
but the peaks are not shaped.
Before the crack growth is completed ($t \le 0.25$), 
\[ \dot{W} > \dot{Q} \approx \dot{E}, \]
indicating that the system energy change
is comparable to the viscous dissipation. 

% Figure 5
%% temporal evolution of energy
% alp_e=0.02,0.05,0.1,1.0, g_0=1
%F5%%%%%%%%%%%%%%%%%%%%%%%%%%%%%
%%%%%%%%%%%%%%%%%%%%%%%%%%%%%%%%
%%%%%%%%%%%%%%%%%%%%%%%%%%%%%%%
% Figure 6
% temporal evolution of energy
%%%%%%%%%%%%%%%%%%%%%%%%%%%%%%%%%%%%%%%%%%

%%%%%%%%%%%%%%%%%%%%%%%%%%%%

\section{Discussion}

As shown in Fig.~\ref{fig:Maxwell-mode3-mu1-gam-0.01-g0-1-alp3-0.1}
-\ref{fig:Maxwell-mode3-g02-1-dene},
increasing $ \tilde{g}_{0}$ results in the crossover
of the mechanical behaviour from ductile,
for which the crack propagation
is inhibited by the increase in $ \mathbf{e} $,
to quasi-brittle with a distinct
crack propagation from the initial crack.
Because
$ \tilde{g}_{0}=\frac{g_{0}}{L /\tau} =\frac{ \tau }{T } $,
distinct crack propagations occur
when the viscoelastic relaxation time $\tau$ is sufficiently
longer than the characteristic time
of the deformation $T$.
This is qualitatively consistent with experiments
in Maxwell-type viscoelastic
liquids~\cite{gladden2007motion}.

The non-dimensionalised equations \eqref{eq:_ue}-\eqref{eq:az_ez_2}
provide a more detailed insight
into the crossover in the deformation behaviour.
First, it is observed that the crack extension, that is, the increase in $z$
ahead of the initial notch, is caused
by the $ (1-z) |\nabla u -\mathbf{e}|^{2}$ term
in the r.h.s. of eq. \eqref{eq:az_ez_2}.

For a very small $\tilde{g}_{0}$, 
Eq. \eqref{eq:ge_ue} means $\nabla u - \mathbf{e} \approx \mathbf{0}$ for $z < 1$
(the viscous strain $ \mathbf{e} $ can follow $\nabla u$
without delay),
thus nullifying the $(1-z) |\nabla u -\mathbf{e}|^{2}$ term.
The deformation cannot drive the crack growth (ductility).

For a large $\tilde{g}_{0}$,
meanwhile,
$\dot{\mathbf{e}} \approx 0$
(i.e. $\mathbf{e}(\mathbf{x},t)
\approx \mathbf{e}(\mathbf{x},0) = \mathbf{0}$)
holds until $| \nabla u | \sim t $ exceeds $\tilde{g}_{0}$
(because $\dot{\mathbf{e}} \propto 1/\tilde{g}_{0} $,
see Eq. \eqref{eq:ge_ue}):
viscoelastic relaxation barely occurs
for a while from the onset of the boundary displacement.
If values of $\tilde{\epsilon}$ and $\tilde{\gamma}$
allow a sudden growth of the initial crack within
the elasticity dominant period ($\mathbf{e}(\mathbf{x},t)
\approx \mathbf{0}$),
the fracture is essentially brittle,
as in the purely elastic model in \cite{takaishi2009phase}.

In the present numerical condition, 
a crossover occurs at approximately $\tilde{g}_{0}= 0.04$;
however, the threshold value should depend
on the relevant parameters of
$ \tilde{\gamma}$ and $ \tilde{\epsilon}$, $ \tilde{\alpha}_z (=\eta/\mu T)$.

\section{Concluding remarks}
In this study, we constructed a gradient flow PFFM
for the mode-III fracture of a Maxwell-type viscoelastic material,
by extending a purely elastic model \cite{takaishi2009phase}.
In our formulation, utilising viscous strain
is paramount.
It allowed us to directly observe where and how viscoelastic
relaxation occurs in the system, as in Figs. \ref{fig:Maxwell-mode3-mu1-gam-0.01-g0-1-alp3-0.1}-\ref{fig:Maxwell-mode3-g0-1-alp3-Eel-Es}.
More importantly,
it provided a systematic method
to recreate a model describing a particular type of elasticity
(e.g. 3d linear elasticity and several hyperelasticities)
into the corresponding viscoelastic PFFM. 
The procedure is straightforward:
(i) Subtract the viscous strain from
the total strain (the symmetrical part of the displacement gradient)
and multiply the elastic constants by $(1-z)^{2}$
in the expression of elastic energy;
(ii) Construct the total energy
by adding the modified elastic energy and
$E_{\rm dam}$ of Eq. \eqref{eq:E_ze}; 
(iii) Derive the gradient-flow system of the
total energy with respect to the deformation field,
viscous strain, and damage phase field $z$.
The applications of the procedure to generic 2d and 3d linear elasticities
have been presented separately in~\cite{takaishi2009phase}. \\
\\
{\bf Acknowledgments} \\
We acknowledge the financial support from JSPS KAKENHI No.~17K05609.
The authors thanks to 
Yasumasa Nishiura
for his encouragement and fruitful discussions.
This work is partially supported by
the Cross-ministerial Strategic Innovation Promotion Program (SIP).

%%%%%%%% \references %%%%%%%%%%


\begin{thebibliography}{10}

\bibitem{ashby2018materials}
Michael~F Ashby, Hugh Shercliff, and David Cebon.
\newblock {\em Materials: engineering, science, processing and design}.
\newblock Butterworth-Heinemann, 2018.

\bibitem{van2013rheology}
Ton Van~Vliet.
\newblock {\em Rheology and fracture mechanics of foods}.
\newblock CRC Press, 2013.

\bibitem{anderson2017fracture}
Ted~L Anderson.
\newblock {\em Fracture mechanics: fundamentals and applications}.
\newblock CRC press, 2017.

\bibitem{gent1994viscoelastic}
AN~Gent, SM~Lai, C~Nah, and CHI Wang.
\newblock Viscoelastic effects in cutting and tearing rubber.
\newblock {\em Rubber Chemistry and Technology}, 67(4):610--618, 1994.

\bibitem{knauss2015review}
Wolfgang~G Knauss.
\newblock A review of fracture in viscoelastic materials.
\newblock {\em International Journal of Fracture}, 196(1-2):99--146, 2015.

\bibitem{creton2016fracture}
Costantino Creton and Matteo Ciccotti.
\newblock Fracture and adhesion of soft materials: a review.
\newblock {\em Reports on Progress in Physics}, 79(4):046601, 2016.

\bibitem{poulain2018damage}
Xavier Poulain, Oscar Lopez-Pamies, and K~Ravi-Chandar.
\newblock Damage in elastomers: healing of internally nucleated cavities and
  micro-cracks.
\newblock {\em Soft Matter}, 14(22):4633--4640, 2018.

\bibitem{schapery1975theory}
Richard~A Schapery.
\newblock A theory of crack initiation and growth in viscoelastic media.
\newblock {\em International Journal of Fracture}, 11(1):141--159, 1975.

\bibitem{hui1992fracture}
Chung-Yuen Hui, Da-Ben Xu, and Edward~J Kramer.
\newblock A fracture model for a weak interface in a viscoelastic material
  (small scale yielding analysis).
\newblock {\em Journal of applied physics}, 72(8):3294--3304, 1992.

\bibitem{de1996soft}
PG~De~Gennes.
\newblock Soft adhesives.
\newblock {\em Langmuir}, 12(19):4497--4500, 1996.

\bibitem{rahulkumar2000cohesive}
P~Rahulkumar, A~Jagota, SJ~Bennison, and S~Saigal.
\newblock Cohesive element modeling of viscoelastic fracture: application to
  peel testing of polymers.
\newblock {\em International Journal of Solids and Structures},
  37(13):1873--1897, 2000.

\bibitem{persson2005crack}
BNJ Persson and EA~Brener.
\newblock Crack propagation in viscoelastic solids.
\newblock {\em Physical Review E}, 71(3):036123, 2005.

\bibitem{onuki2002phase}
Akira Onuki.
\newblock {\em Phase transition dynamics}.
\newblock Cambridge University Press, 2002.

\bibitem{aranson2000continuum}
IS~Aranson, VA~Kalatsky, and VM~Vinokur.
\newblock Continuum field description of crack propagation.
\newblock {\em Physical review letters}, 85(1):118, 2000.

\bibitem{bourdin2000numerical}
Blaise Bourdin, Gilles~A Francfort, and Jean-Jacques Marigo.
\newblock Numerical experiments in revisited brittle fracture.
\newblock {\em Journal of the Mechanics and Physics of Solids}, 48(4):797--826,
  2000.

\bibitem{karma2001phase}
Alain Karma, David~A Kessler, and Herbert Levine.
\newblock Phase-field model of mode iii dynamic fracture.
\newblock {\em Physical Review Letters}, 87(4):045501, 2001.

\bibitem{ambati2015phase}
Marreddy Ambati, Tymofiy Gerasimov, and Laura De~Lorenzis.
\newblock Phase-field modeling of ductile fracture.
\newblock {\em Computational Mechanics}, 55(5):1017--1040, 2015.

\bibitem{miehe2015phase}
Christian Miehe, M~Hofacker, L-M Sch{\"a}nzel, and Fadi Aldakheel.
\newblock Phase field modeling of fracture in multi-physics problems. part ii.
  coupled brittle-to-ductile failure criteria and crack propagation in
  thermo-elastic--plastic solids.
\newblock {\em Computer Methods in Applied Mechanics and Engineering},
  294:486--522, 2015.

\bibitem{carrara2017consistent}
P~Carrara and L~De~Lorenzis.
\newblock Consistent identification of the interfacial transition zone in
  simulated cement microstructures.
\newblock {\em Cement and Concrete Composites}, 80:224--234, 2017.

\bibitem{chukwudozie2019variational}
Chukwudi Chukwudozie, Blaise Bourdin, and Keita Yoshioka.
\newblock A variational phase-field model for hydraulic fracturing in porous
  media.
\newblock {\em Computer Methods in Applied Mechanics and Engineering},
  347:957--982, 2019.

\bibitem{miehe2014phase}
Christian Miehe and Lisa-Marie Sch{\"a}nzel.
\newblock Phase field modeling of fracture in rubbery polymers. part i: Finite
  elasticity coupled with brittle failure.
\newblock {\em Journal of the Mechanics and Physics of Solids}, 65:93--113,
  2014.

\bibitem{shen2019fracture}
Rilin Shen, Haim Waisman, and Licheng Guo.
\newblock Fracture of viscoelastic solids modeled with a modified phase field
  method.
\newblock {\em Computer Methods in Applied Mechanics and Engineering},
  346:862--890, 2019.

\bibitem{ambrosio1992approximation}
L~Ambrosio and VM~Tortorelli.
\newblock On the approximation of functionals depending on jumps by quadratic,
  elliptic functionals.
\newblock {\em Boll. Un. Mat. Ital}, 6:105--123, 1992.

\bibitem{takaishi2009phase}
Takeshi Takaishi and Masato Kimura.
\newblock Phase field model for mode iii crack growth in two dimensional
  elasticity.
\newblock {\em Kybernetika}, 45(4):605--614, 2009.

\bibitem{takaishi2009numerical}
Takeshi Takaishi.
\newblock Numerical simulations of a phase field model for mode iii crack
  growth (theory, continuum mechanics focusing singularities, $\langle$ special
  issue $\rangle$ joint symposium of jsiam activity groups 2009).
\newblock {\em Transactions of the Japan Society for industrial and Applied
  Mathematics}, 19(3):351--369, 2009.

\bibitem{kimura2019gradient}
Masato Kimura, Hirofumi Notsu, Yoshimi Tanaka, and Hiroki Yamamoto.
\newblock The gradient flow structure of an extended maxwell viscoelastic model
  and a structure-preserving finite element scheme.
\newblock {\em Journal of Scientific Computing}, pages 1--21, 2019.

\bibitem{gladden2007motion}
Joseph~R Gladden and Andrew Belmonte.
\newblock Motion of a viscoelastic micellar fluid around a cylinder: flow and
  fracture.
\newblock {\em Physical review letters}, 98(22):224501, 2007.

\bibitem{tabuteau2011propagation}
Herv{\'e} Tabuteau, Serge Mora, Matteo Ciccotti, Chung-Yuen Hui, and Christian
  Ligoure.
\newblock Propagation of a brittle fracture in a viscoelastic fluid.
\newblock {\em Soft Matter}, 7(19):9474--9483, 2011.

\bibitem{huang2017polymer}
Qian Huang and Ole Hassager.
\newblock Polymer liquids fracture like solids.
\newblock {\em Soft Matter}, 13(19):3470--3474, 2017.

\bibitem{hecht2012new}
Fr{\'e}d{\'e}ric Hecht.
\newblock New development in freefem++.
\newblock {\em Journal of numerical mathematics}, 20(3-4):251--266, 2012.

\end{thebibliography}
\end{document}